# Tests & Calibration on Ultra Violet Imaging Telescope (UVIT)

Amit Kumar[1], S. K. Ghosh[2], P.U.Kamath[1], Joe Postma[4], S. Kathiravan[1], P.K.Mahesh[1], Nagbhushana.S [1], K.H.Navalgund[3], N Rajkumar[1] M.N.Rao[1], K.S.Sarma[3], S.Sriram[1], C. S. Stalin[1], S.N.Tandon[1,5]

1. Indian Institute of Astrophysics, Bangalore
2. National Centre for Radio Astrophysics, TIFR, Pune
3. ISRO Satellite Centre, Bangalore
4. Department of Physics and Astronomy, University of Calgary, Calgary, AB, Canada
5. Inter University Centre for Astronomy and Astrophysics, Pune

## ABSTRACT

Ultra Violet Imaging Telescope on ASTROSAT Satellite mission is a suite of Far Ultra Violet (FUV; 130 – 180 nm), Near Ultra Violet (NUV; 200 – 300 nm) and Visible band (VIS; 320–550nm) imagers. ASTROSAT is a first multi wavelength mission of INDIA. UVIT will image the selected regions of the sky simultaneously in three channels & observe young stars, galaxies, bright UV Sources. FOV in each of the 3 channels is ~ 28 arc-minute. Targeted angular resolution in the resulting UV images is better than 1.8 arc-second (better than 2.0 arc-second for the visible channel). Two identical co-aligned telescopes (T1, T2) of Ritchey-Chretien configuration (Primary mirror of ~375 mm diameter) collect the celestial radiation and feed to the detector system via a selectable filter on a filter wheel mechanism; gratings are available in the filter wheels of FUV and NUV channels for slit-less low resolution spectroscopy. The detector system for each of the 3 channels is generically identical. One telescope images in the FUV channel, and other images in NUV and VIS channels. One time open-able mechanical cover on each telescope also works as Sun-shield after deployment.

We will present the optical tests and calibrations done on the two telescopes. Results on vibrations test and thermo-vacuum tests on the engineering model will also be presented.

## 1. INTRODUCTION

The Ultra Violet Imaging Telescope (UVIT) is one of the five instruments to go on the Indian multi-wavelength mission ASTROSAT. This instrument makes high resolution (FWHM < 1.8") images in a field of ~ 28' simultaneously in three channels in ultraviolet and visible: VIS (320-550 nm), NUV (200-300 nm), and FUV (130-180 nm); the other four instruments observe in soft/hard X-rays. UVIT is configured as twin telescopes – one of these images in VIS & NUV and the other in FUV. Preparation of such a payload involves testing of the performance of the components as well as that of sub-assemblies and the full assembly. In addition environmental tests are performed to ensure survival through vibrations of the launch and thermal variations in the orbit. In this paper we shall present results of the tests done on the individual telescopes of the flight model, and the environmental tests done on the engineering model.

*amits@iiap.res.in; phone 91 80 22541253; fax 91 80 25534043; iiap.res.in

## 2. UVIT INSTRUMENT OVERVIEW

UVIT is configured as twin telescopes; each telescope is a Ritchey Chertian (RC) configuration; having an aperture of ~375mm and a focal length of ~4750mm. One of the two telescopes is fully dedicated to FUV (130-180 nm) channel, while the other telescope is used for NUV (200-300 nm) and VIS (320-550 nm) channels. Each of the telescopes is having Field of View of ~28', with a FWHM better than 1.8" – this can be compared with the values of 4.5" to 5" for GALEX[2]. To select a particular narrow wavelength band, for any of the three channels, a set of filters has been provided on wheel. The wheels for the ultraviolet channels also carry a grating to provide low resolution slitless spectroscopy. Adequate baffling has been provided to avoid any stray light due to bright off-field sources. An attenuation of $10^9$ is obtained for sources at 45º from the axis. One time openable door on each telescope provides sun shade as long as Sun is at > 45º from the axis. Intensified imagers are used as photon counting detectors (see reference 1).

Main structure of UVIT telescopes which holds all the optical components like telescope tubes, telescope rings, detector mounting brackets, spider rings etc is made of Invar36, and the other parts like are made of aluminum alloy. A cone-like structure of titanium material is used to interface the two telescopes to the S/C. Active thermal control is used to minimize defocus etc. in the optics due to temperature variations in the orbit. (see reference 3).

## 3. OPTOMECHANICAL ALIGNMENT

### 3.1 Alignment of Individual Telescopes

Optomechanical alignment of flight model individual telescope was carried out by using alignment telescope cum auto collimator (for coarse alignment) and ZYGO interferometer (fine alignment). Details of optical layout of each telescope are given in reference 3. Following table shows the required parameters and obtained parameters of the individual telescope alignment.

Table 1 Opto-mechanical alignment Parameters for Individual Telescope

| S.No. | Alignment Objectives | Requirements | Obtained |
|---|---|---|---|
| 1. | Primary Secondary inter Separation | Correct to 0.1 mm | < 0.1 mm |
| 2. | Coma on wavefront under zero gravity | < 0.1 λ | < 0.05 λ |
| 3. | Astig on wavefront under zero gravity | < 0.1 λ | < 0.07 λ |
| 4. | Alignment of Detector centre to optical axis | < 30 arc sec | < 20 arc sec |
| 5. | Alignment of Detector rows/Coloums wrt telescope Pitch/Yaw axis with in | < 60 arc sec | < 30 arc sec |
| 6. | Relative alignemnt of NUV to VIS Detectors center | < 30 arc sec | < 20 arc sec |
| 7. | Detector position wrt telescope focus | Within 100 microns | < 100 microns |

Unit telescope is aligned with interferometer to meet the above optical alignment requirement. ZYGO interferometer is used in autocollimation test configuration mode with reference flat (RMS λ/100 -Optical Surface Ltd. UK) as shown in figure 1.

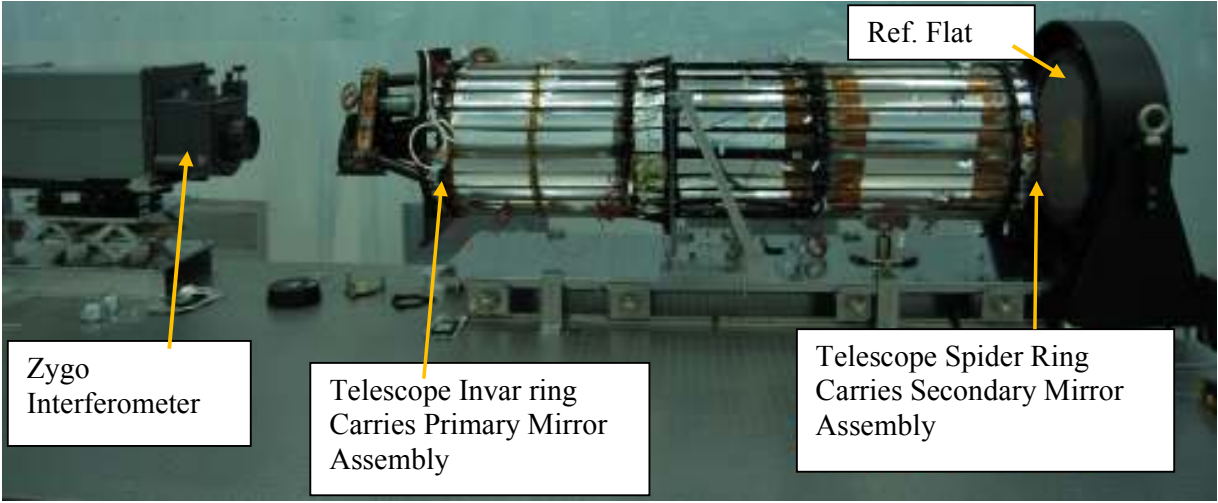

Figure 1 Interferometric alignment test setup on optical bench in 100 Class Clean room

### 3.2 Alignments of the Two Telescopes

In order to have an overlap of the fields in the three channels, it is required that optical-axes of the two telescopes be well aligned relative to each other, and centres of the detectors be on the corresponding optical axis. Further, axes of the two telescopes should be aligned to axes of the S/C. The optical axis of each telescope, defined as axis of the primary mirror, is transferred to three locations: i) rear surface of the primary mirror, ii) rear surface of the secondary mirror, and iii) an alignment cube mounted on the bracket which supports the detector. Parallelism between the two telescopes is measured by measuring the angle between rear surfaces of the two secondary mirrors (if need be the other two references can be used). An additional alignment cube is provided on the satellite-interface-adaptor for alignment with the S/C. The requirements on the alignments are listed in Table 2.

Table 2 Opto-mechanical alignment parameters for the two telescopes

| S.No. | Alignment Objective | Requirement |
|---|---|---|
| 1. | Co-alignment of centers of three detectors | < 30 arc second |
| 2. | VIS/NUV telescope optical axis to Alignment cube on Ti cone( satelite interface adaptor) | < 60 arc second |
| 3. | Parallism between the two telescopes | < 60 arc second |

## 4. OPTICAL TESTS AND CALIBRATIONS

The three detector systems were individually tested for uniformity of sensitivity (see reference 4) and quantum efficiency (see reference 5), and filters were tested for transmission and its uniformity (see reference 6). After assembly each of the two telescopes are being tested for PSF in the detectors and for the best focus point.

## 5. ENVIRONMENTAL TESTS ON ENGINEERING MODEL

### 5.1 Electromagnetic Interference & Electromagnetic Compatibility (EMI-EMC) Test

The detector system of paylaod is made of many parts, and these parts sit at different locations. The main Electronic Unit "EU", which is the link to spacecraft (S/C), sits on the S/C bus, while 3 detectors & 3 high voltage supplies sit in the focal volume of payload. Thus, analogue video signals from the detectors are carried over long cables. In order to check that any noise in the video signals due to EMI is within acceptable limits, an EMI-EMC test is conducted, at ISAC-ISRO, by simulating configuration of the ground plane in the S/C. The simulated configuration is shown in Figure 2.

The main aim of conducting EMI/ EMC test is to check that the read noise in the video signals is acceptable, and the communications to and from the detector system are reliable. EM payload has qualified the EMI-EMC test in the above configuration except a little deviation for radiated susceptibility test at frequency bands of 33MHz to 34.8MHz and 59.5MHz to 63MHz. Observed maximum read noise (acceptable value is $\leq 30$)[7] in frequency band 33MHz to 34.8MHz is 31.28 and for frequency band 59.5MHz to 63MHz is 88.48; this large noise of 88.88 could increase FWHM of the PSF from 1.8" to 1.9".

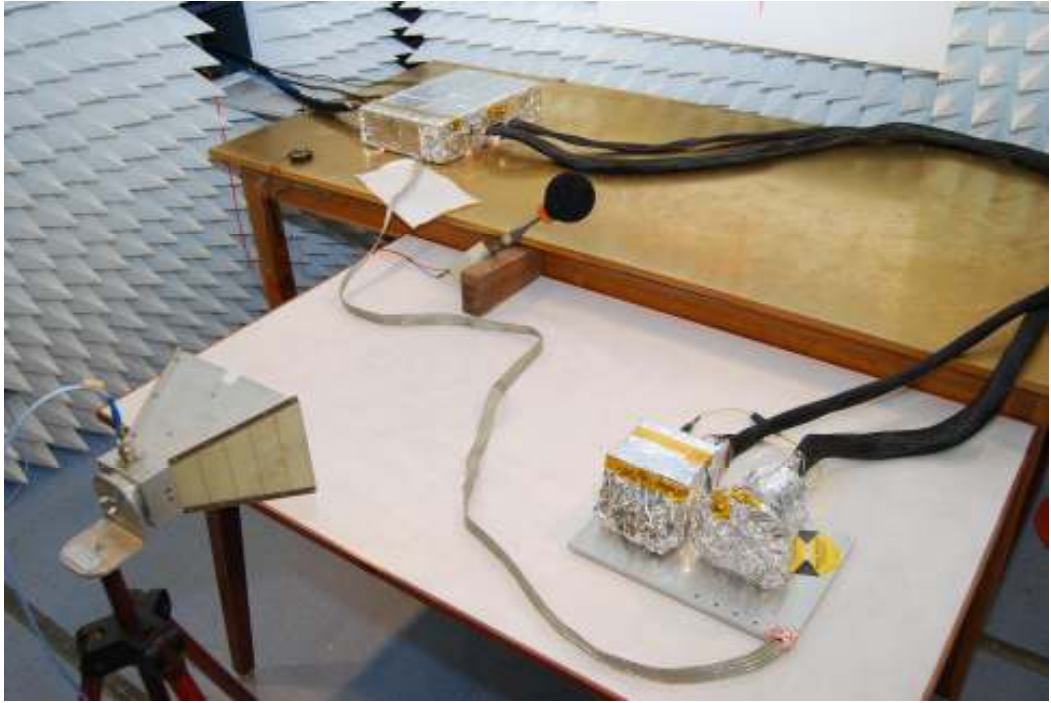

Figure 2 EM UVIT Detector System Test Setup at ISRO EMI-EMC test facility. Electronics Unit is sitting (back side) on ground plane-1 and detector and high voltage unit (front side) on ground plane-2.

**5.2 Vibration Test**

Vibration tests were performed on the engineering model of the payload to verify the dynamic characteristics of the payload, to correlate and update the mathematical model, to demonstrate the mechanical integrity of the payload and qualify the payload to the launch loads. Prototype test philosophy was adopted to qualify the payload. Both Sine vibration and Random vibration tests were conducted in all the three mutual perpendicular axes of the payload. Extensive acceleration response monitoring (see figure 3) was done for the payload to derive the dynamic characteristics and also to safeguard the critical subsystems. A vibration test fixture was designed (see figure 4) to simulate the payload mechanical mounting interface. The fixture survey test was carried out to understand its dynamic behavior. The tests were carried out on 16 Ton capacity electro-dynamic shaker. Where the mechanical configuration includes more than one similar components, e.g. detectors and high voltage units/ mirrors/ structures of the two telescopes, it was decided to use only one real component and replace the others by dummies of equivalent rigidity and mass/volume. The payload vibration test setup is shown in figure 5.

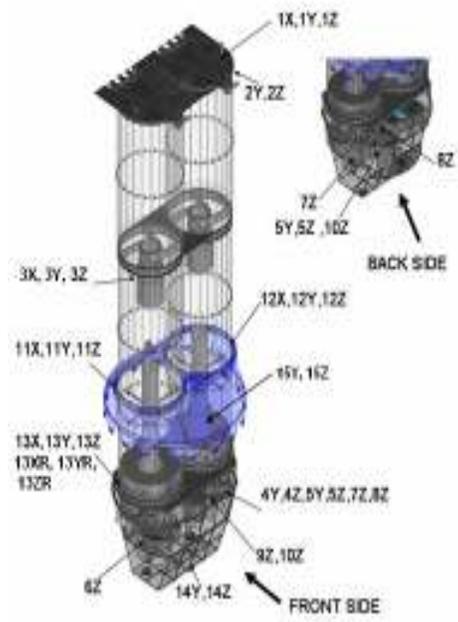 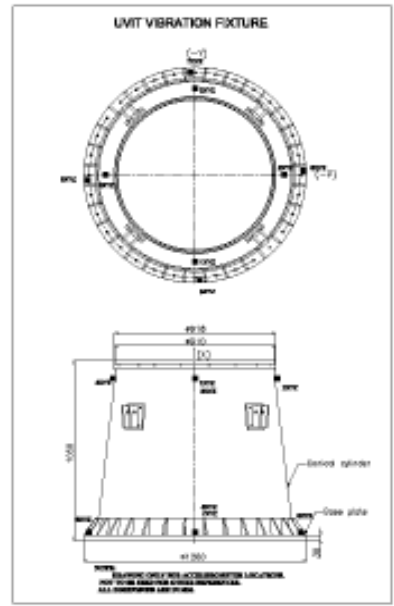

Figure 3 Response Monitoring Locations    Figure 4 Vibration Test Fixture

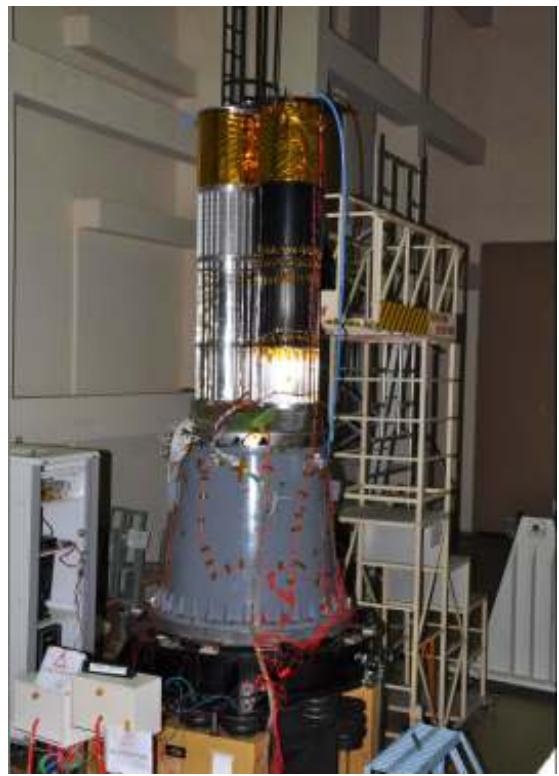

Figure 5 Vibration test setup of EM UVIT mounted on vibration test fixture at ISRO vibration test facility

**5.2.1 Test Specifications:**

The recommended vibration test specifications (base input to the payload interface) used for the tests are as given below.

Table 3 Proto-flight Level Sine Test Specifications (X axis)

| Freq. Range | Levels | Tol.(dB) | Abort (dB) | Sweep rate |
|---|---|---|---|---|
| 5 - 16 Hz | 9.7 mm | +/- 2 | +4 / -6 | |
| 16 - 50 Hz | 10 g | +/- 2 | +4 / -6 | 2 Oct./ minute |
| 50-80 Hz | 6 g | +/- 2 | +4 / -6 | |
| 80 -100 Hz | 3.5 g | +/- 2 | +4 / -6 | |

Table 4 Proto-flight Level Sine Tests (Y and Z axis)

| Freq. Range | Levels | Tol.(dB) | Abort(dB) | Sweep rate |
|---|---|---|---|---|
| 5 - 11 Hz | 10.3mm | +/- 2 | +4 / -6 | |
| 11 - 60 Hz | 5 g | +/- 2 | +6 / -6* | 2 Oct./ minute |
| 60 - 100 Hz | 3 g | +/- 2 | +4 / -6* | |

Table 5 Proto-flight Level Random Test Specifications for All Three Axis of Payload

| Frequency (Hz) | Qualification Level PSD* |
|---|---|
| 20 -100 | + 3dB/Octave |
| 100 – 700 | 0.05 $g^2$ / Hz |
| 700 – 2000 | -3dB/Octave |
| Overall g RMS | 8.3 g (rms) |
| Duration | 60 seconds |

* PSD - Power Spectral Density

**Notching Philosophy:** Response limiting was adopted during the tests to safeguard the critical subsystems of the payload against the excessive loading due to impedance mismatch due to hard mounting of the payload with the shaker. The payload global mode acceleration response was limited to 15g at the payload CG (which is equivalent to the recommended quasi-static loads for the payload) while the responses of local resonances of individual subsystems were limited to the levels to which they were qualified during their unit level tests.

The qualification tests on the payload were successfully completed. The mechanical integrity of the payload was established by pre and post signature test data. The mechanical alignment stability before and after the qualification tests were correlating. The functional tests of the payload were satisfactory.

## 5.3 Thermo-Vacuum Test:

Thermo-vacuum (TVAC) test was done on Engineering Model (EM) of payload to demonstrate that the payload system remains operational and not damaged during and after exposures to qualification levels of thermo vacuum conditions specified by ETLS. This test had given us an experience and modalities that will be followed for achieving the cleanliness and to minimize the contamination levels for flight model payload/ spacecraft tests. This was an active test; detector, high voltage supplies, filter wheel motor, electronics unit, filter wheel drive electronics were operational during the test. Following temperature profile as shown in figure 6 was followed to qualify payload for TVAC test.

The test was carried out in ISRO TVAC test facility. During the test maintained vacuum level was better than $1\times10^{-6}$ mbar. Chamber shroud was maintained at -80°C and UVIT instrument temp was maintained as shown in figure 6 through 123 IR lamps. For temperature monitoring combination of thermistors and thermo-couples were used.

To remove any possible contaminants from TVAC test chamber, empty chamber was baked at +60°C. Baking continued till TCQM showed rate $< 8.71\times10^{-12}$/gm/cm$^2$/sec, and a rate of change in rate of $<3.03\times10^{-16}$/gm/cm$^2$/sec$^2$ for 8 hrs. Transmission loss for MgF$_2$ witness samples was 1-1.50% during empty chamber baking at wavelength band 130nm-160nm. After getting the satisfactory results from TCQM and MgF$_2$ witness samples, paylaod instrument was placed inside the TVAC chamber. During TVAC test also, for contamination monitoring TCQM and MgF$_2$ witness samples were used. TVAC test setup is shown in figure 7.

After the completion of the test data transmission loss through MgF$_2$ witness samples were analyzed and it was found that the transmission loss of MgF$_2$ window is 2-2.5% at 130-160nm wavelength band.

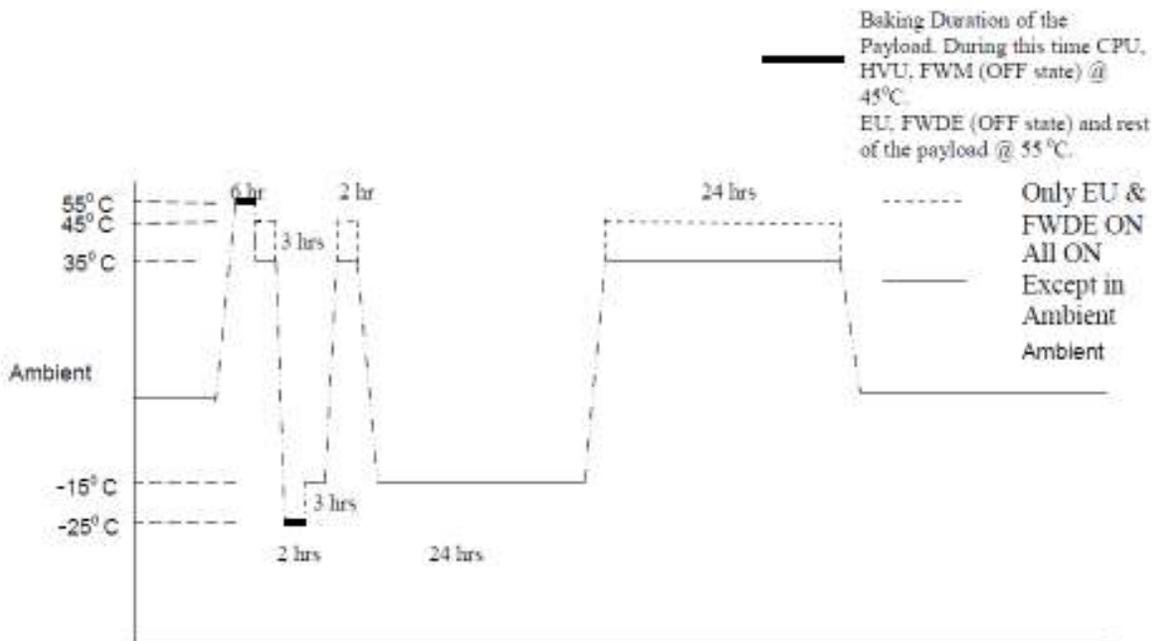

Figure 6 EM UVITDS TVAC test profile (qualification level)

The performance of detector, high voltage supply, filter wheel motor, electronics unit and filter wheel drive electronics were being checked at each temperature plateau. Post TVAC also the performance of all the electrical unit was checked and found normal. Payload qualified the TVAC test successfully.

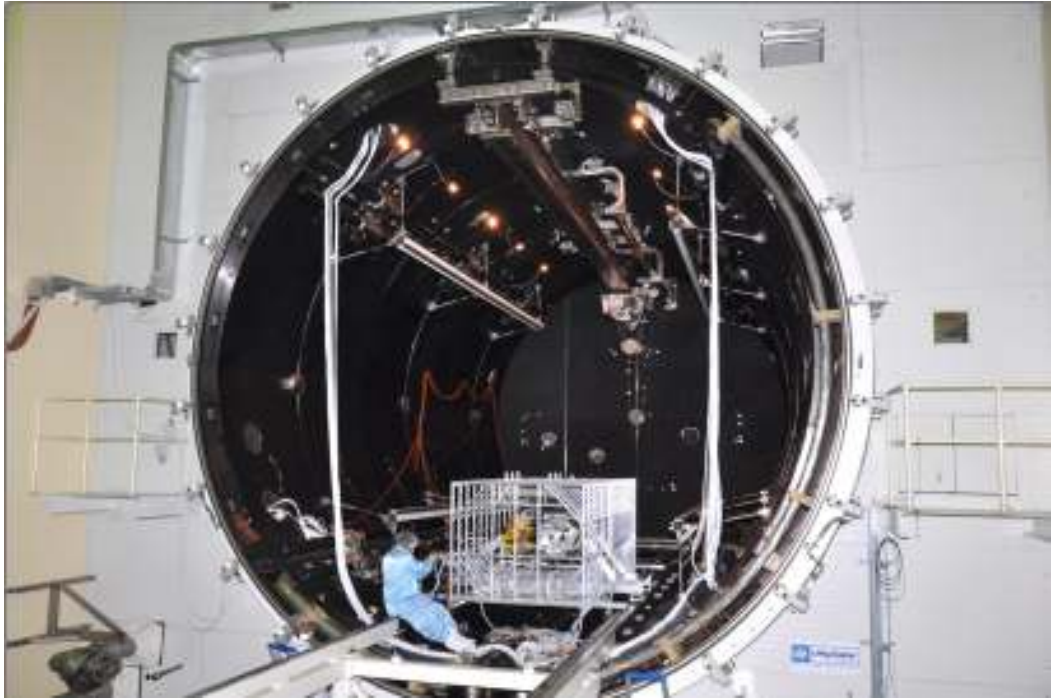

Figure 7 TVAC test Setup of EM UVIT at ISRO thermo-vacuum test facility. Payload is enclosed in a fixture having Infra Red Lamps to heat the payload. This Fixture is mounted and engaged with the rails of TVAC chamber.

## 6. CONCLUSION

EM UVIT has successfully qualified all the environmental tests required to go ahead with the integration of flight model in 2011. Currently; flight model individual telescopes are ready and undergoing optical tests. After completion of the optical tests both the telescopes will be integrated together in satellite adapter and final integration of instrument will be completed very soon. The launch of the instrument with Astrosat spacecraft is expected by 2013.

## ACKNOWLEDGMENTS

The UVIT project is collaboration between the following institutes from India: Indian Institute of Astrophysics (IIA), Bengaluru, Inter University Centre for Astronomy and Astrophysics (IUCAA), Pune, and National Centre for Radioastrophysics (NCRA) (TIFR), Pune, and the Canadian Space Agency (CSA). Many departments and environmental test facilities (EMI-EMC, Vibration and Thermo-vacuum) from ISAC, ISRO, Bengaluru have provided direct support in testing and qualification of the payload. We give our sincere thanks to B.S.Nataraju and M.N.Karthikeyan for their valuable support and consultation. We also thanks to all technicians, trainees and other supporting staff, involved in this for their untiring support.

# REFERENCES

[1] Hutchings, J.B., Postma J, Asquin D, Leahy D, "Photon Event Centroiding with UV Photon-counting Detectors", PASP, 119-1152 (2007).

[2] Oswald Siegmund et al, "The GALEX Mission and Detectors", Proceedings of SPIE, Vol. 5488, 13-24 (2004).

[3] Amit Kumar et al., "Ultra Violet Imaging Telescope (UVIT) on ASTROSAT", Proceedings of SPIE (2012), Vol. 8443.

[4] J. Postma, J. B. Hutchings, D. Leahy, "Calibration and Performance of the Photon-counting Detectors for the Ultra Violet Imaging Telescope (UVIT) of the Astrosat Observatory", PASP, 123: 833-843 (2011).

[5] Stalin, C.S., Sriram, S., Amit Kumar, "Determination of Quantum Efficiency of UVIT Flight Model Detectors", Indian Institute of Astrophysics, Bangalore, http://www.iiap.res.in/files/corrected_detector_1.pdf, (2010).

[6] K. Sankarsubramanian et al, "Calibration of the FM filters for UVIT", UVIT-CDR-00-006, Jun 2011.

[7] Asquin Don, "UVIT CDR Environmental Testing", UVIT-CDR presentation, Oct, 2008.